\documentclass[]{article}

\usepackage[utf8]{inputenc}
\usepackage{lineno,hyperref}
\usepackage{longtable}
\usepackage{graphicx}
\usepackage{graphicx}
\usepackage{amsmath} 
\usepackage{xcolor}

\title{FBDNN: Filter Banks and Deep Neural Networks for Portable and Fast Brain-Computer Interfaces}
\author{Pedro R. A. S. Bassi\thanks{Department of Computer Engineering and Industrial Automation, School of Electrical and Computer Engineering, University of Campinas - UNICAMP. 13083-970, Campinas, SP, Brazil. E-Mail: p157007@dac.unicamp.br.} and Romis Attux}
\date{}

\providecommand{\keywords}[1]
{
  \small	
  \textbf{\textit{Keywords---}} #1
}

\begin{document}

\maketitle

\begin{abstract}

\emph{Objective.}
To propose novel SSVEP classification methodologies using deep neural networks (DNNs) and improve performances in single-channel and user-independent brain-computer interfaces (BCIs) with small data lengths.
\emph{Approach.}
We propose the utilization of filter banks (creating sub-band components of the EEG signal) in conjunction with DNNs. In this context, we created three different models: a recurrent neural network (FBRNN) analyzing the time domain, a 2D convolutional neural network (FBCNN-2D) processing complex spectrum features and a 3D convolutional neural network (FBCNN-3D) analyzing complex spectrograms, which we introduce in this study as possible input for SSVEP classification.
We tested our neural networks on three open datasets and conceived them so as not to require calibration from the final user, simulating a user-independent BCI.
\emph{Results.}
The DNNs with the filter banks surpassed the accuracy of similar networks without this preprocessing step by considerable margins, and they outperformed common SSVEP classification methods (SVM and FBCCA) by even higher margins. 
\emph{Conclusion and significance.}
Filter banks allow different types of deep neural networks to more efficiently analyze the harmonic components of SSVEP. Complex spectrograms carry more information than complex spectrum features and the magnitude spectrum, allowing the FBCNN-3D to surpass the other CNNs. The performances obtained in the challenging classification problems indicates a strong potential for the construction of portable, economical, fast and low-latency BCIs.

\end{abstract}\hspace{10pt}

\newpage

\section{Introduction}

Recent studies have addressed the creation and improvement of small and portable brain-computer interfaces (BCI) \cite{SingleChannelCNN}, \cite{BSPCWork}, \cite{singleChenel2}, which have the potential of being wearable and less expensive, as well as of allowing a simple self-application.  Having this in mind, this work proposes new classification methodologies for high-speed and low latency (based on small data lengths, i.e., analyzing signals with short duration/sample length) single-channel SSVEP BCIs. We introduce the utilization of filter banks as a pre-processing step for recurrent neural networks (RNNs, analyzing the time domain) and for CNNs analyzing the complex frequency domain or complex spectrograms, a novel type of input in the field of SSVEP classification. 

Steady state visually evoked potentials (SSVEPs) establish a paradigm in BCI with very interesting properties, like noninvasiveness and a relatively high signal-to-noise ratio (SNR) \cite{galloway1990}. In this work, we utilize a single-channel approach, analyzing only the electroencephalography (EEG) signal from the Oz electrode, to simulate portable, less expensive and practical devices. 

Many methods have been used to classify SSVEPs, being canonical correlation analysis (CCA) \cite{CCA} a widespread instance. A filter bank, composed of band-pass filters, was introduced in \cite{FBCCA} to allow CCA to better obtain information from the SSVEP harmonics. The authors of \cite{FBCCA} called this technique Filter Bank Canonical Correlation Analysis (FBCCA), and, in their study, it successfully improved the BCI performance in relation to CCA.

Other works used machine learning (ML) methods to classify SSVEPs. Support vector machines (SVMs \cite{SVmOriginal}) are commonly employed, since they are relatively easy to train and perform well in terms of accuracy. An example is \cite{SVMExComparison}, which used a SVM based BCI to control a RF car and compared different types of SVM kernels. Another study \cite{SVMEx} applied a SVM to classify a 14-channel BCI.

In the last few years, more works have used convolutional neural networks (CNNs) as SSVEP classifiers, generally analyzing the EEG signals in the frequency domain (using fast Fourier transform, or FFT \cite{fft}). For example, in \cite{MultiChannelCNN} the authors utilized data from 8 electrodes, in the frequency domain, to create the 8 rows of a matrix, which fed a two-dimensional convolutional neural network. Their network performed better than CCA, with 94\% accuracy, using a data length of 2 s and they employed their BCI to control an exoskeleton in an ambulatory environment. 

Recently, more studies are using deep neural networks for the classification of single-channel SSVEP. As an example, a work \cite{SingleChannelCNN} created a portable BCI that used a light 3D-printed helmet with a single electrode. They employed a one-dimensional deep convolutional neural network to analyze the signals in the frequency domain and their network performance surpassed CCA when the subjects were moving. Another study \cite{singleChenel2} classified 1-channel SSVEP with a recurrent neural network (RNN, based on long-short term memory, LSTM \cite{LSTM}), analyzing the signal in the time domain, and compared it with a CNN, analyzing the frequency domain. They used the BCI to successfully control a quadcopter and observed better performance with the RNN (achieving 92.9\% accuracy with the RNN and 56.6\% using the CNN, with a 0.5 s data length).

The utilization of complex spectrum features and CNNs to classify SSVEP was recently introduced by \cite{ComplexSSVEP}. The study applied FFT to the EEG signals, and, for each channel, it concatenated the real and imaginary parts of the operation's output, forming a single vector. Afterwards, it stacked the vectors for each electrode, forming a matrix that served as the CNN input. Unlike the aforementioned studies, \cite{ComplexSSVEP} created a user-independent BCI. Using 1 s data length and 3 electrodes, they observed 81.6\% mean accuracy with the complex spectrum features, against 70.5\% with the common magnitude features. The idea behind using complex spectrum features is to preserve the magnitude and phase information of the signal. Considering that different visual stimuli in SSVEP generally have different phases, the usefulness of the phase information becomes apparent.

In a previous work \cite{BSPCWork} we used CNNs to classify single-channel SSVEP and we noted that the DNNs surpassed FBCCA when using a single electrode and a high-speed BCI (data length of 0.5 s). In that work, we utilized spectrograms (magnitude) of the EEG signals, and obtained a mean accuracy of 81.8\%.

In the study presented here, we hypothesized that a filter bank would allow deep neural networks to better extract information from the different harmonics in SSVEP, as it allowed CCA. Therefore, we propose 3 DNN architectures that utilize this concept: FBCNN-2D, FBCNN-3D and FBRNN. We will refer to the 3 models as FBDNNs. Inspired by the results shown in \cite{ComplexSSVEP}, the FBCNN-2D is a two-dimensional convolutional neural network that analyzes complex spectrum features. However, we introduce a new strategy to create the CNN input matrix from the FFT real and imaginary parts. The FBCNN-3D is a three-dimensional CNN created to analyze a novel type of input, a complex spectrogram. Its advantage is to preserve, besides the information related to phase and magnitude (already present in the complex spectrum features), more information related to the time domain. Finally, the FBRNN is a recurrent neural network, based on 1-D convolutions and LSTM layers, which analyzes the EEG signals in the time domain. All the proposed DNNs utilize a filter bank as a preprocessing step. Therefore, we will also investigate if filter banks can be a successful general preprocessing stage for different DNN architectures in the field of SSVEP classification.

%Briefly, the FBCNN-2D has a filter bank, with band-pass filters, producing sub-band components of the EEG signal and it utilizes the fast Fourier transform (FFT) to convert the sub-band components to the frequency domain, creating vectors that contain the FFT's real and imaginary parts. Afterwards, the vectors are stacked, creating a matrix that a two-dimensional deep CNN analyzes. 

%The FBCNN-3D also has a filter bank producing sub-band components of the EEG signal, but it does not use the FFT. Instead, it uses a short-time Fourier transform (STFT \cite{bookSTFT}) to create complex spectrograms of the signals, which are matrices containing the STFT's result real and imaginary parts. The spectrograms are stacked, generating a 3D tensor, which we classify using a three-dimensional deep CNN.

%At last, the FBRNN begins with the same filter bank, and we stack the generated sub-band components, creating a matrix (showing signals in the time domain). A DNN composed of 1-D convolutional layers followed by LSTM units classifies this input.

We also implemented other SSVEP classification methods, which will allow us to evaluate the benefits of the filter banks and of our proposed classification schemes. These methods are: FBCCA, Random Forest, SVM and DNNs without the filter banks. To simulate BCIs that do not require calibration with the final user (user-independent/cross-subject approach), we never used the test subject data for training or validation. We conducted tests on three open datasets, Benchmark \cite{dataset}, BETA \cite{BETA} and ``Portable"\cite{portable}. The first two use medical-grade, 64-channel whole head electroencephalography, but we only analyze the Oz electrode, to simulate a single-channel device. The third one has data from a portable interface, with 8 electrodes, from which we also consider only one (Oz).

In our search, we could find only a single SSVEP study that also combined filter banks and neural networks \cite{OtherFBPaper}. However, their presented approach is fundamentally different from what we propose here: they utilize the filter banks with a CNN in the time domain and analyze a subject-dependent BCI with 9 electrodes. Their network obtained very strong performances and  they claimed to have the highest reported information transfer rates in two public SSVEP datasets \cite{OtherFBPaper}. 

To the extent of the author's knowledge, this is the first study in the field of SSVEP classification to employ filter banks combined with RNNs and with CNNs analyzing complex spectrum features. Furthermore, it is also the first study to utilize complex spectrograms or 3D CNNs for SSVEP classification.

\section{Methods}

\subsection{SSVEP-based BCIs}

An SSVEP-based BCI has a visual interface (e.g. a computer monitor) showing flickering visual stimuli. When the users focus their attention on a stimulus, electrical activity (associated with the SSVEPs) in the frequency of stimulation and its harmonics is generated in their brain, mostly in the visual cortex, and it can be captured by electroencephalography. A classifier, like those presented in this study, analyzes the EEG signals, and it must identify the frequency of the visual stimulus that originated the SSVEP. When the users focus their attention on different visual stimuli, with different frequencies, they send different commands to the BCI \cite{beverina}. 

Generally, the interface analyzes signals from a combination of electrodes, but there is a significant interest in enhancing BCI portability/wearability \cite{SingleChannelCNN}. The utilization of a single electrode can reduce the device size, cost and make it easier to wear. In this work, we will focus on this kind of BCI, hence we will utilize data exclusively from the Oz electrode, which is placed above the visual cortex \cite{bassi_rampazzo_attux_2019}.

It is important to remark that we will consider that the BCI does not require calibration for the final user, i.e., we will create a user-independent system. Therefore, we will not use data from the test subject in the training and validation datasets.

The data length is an important characteristic of a BCI. It defines the signal duration that is processed by the system to produce a classification, and smaller lengths create models that are faster and have lower latency. We will analyze  signals of 0.5 s, which is on the lower boundary of the values that still allow enough frequency resolution in FFT/STFT to distinguish between the target frequencies that we will classify in the Benchmark \cite{dataset} and BETA \cite{BETA} datasets (12 and 15 Hz). For more similar stimulus frequencies (e.g., 9.25 Hz and 9.75 Hz, two stimuli that we will consider in the Portable \cite{portable} dataset) this resolution will only allow differentiation in higher SSVEP harmonics.

The utilization of a single electrode, a small data length and a user-independent BCI are approaches that make the classification problem more difficult and tend to reduce accuracy \cite{singleChenel2}, \cite{SingleChannelCNN}, \cite{ComplexSSVEP}. Therefore, one of our objectives is to propose DNNs and preprocessing methodologies that improve performance in this challenging scenario. 

\subsection{Datasets}

We utilized the open dataset called Benchmark \cite{dataset}, assembled to evaluate a virtual keyboard, composed of 40 flickering visual stimuli, shown in a computer monitor. Different stimuli had different frequencies, ranging from 8 to 15.8 Hz (with 0.2 Hz intervals between frequencies), and different phases (adjacent flickers had a 0.5 $\pi$ phase difference). When the subjects focused their gaze and attention in a stimulus, they sent a command (letter or symbol) to the BCI. The authors of \cite{dataset} recorded the signals with 64-channel whole head electroencephalography (EEG) and down-sampled them from 1000 Hz to 250 Hz. They also utilized a notch filter to remove the common 50 Hz power line noise. The study had 35 subjects, who observed the visual flickers in 6 blocks of 40 trials (1 trial for each flicker, per block), and each trial lasted 5 seconds. 8 of the subjects were experienced in using BCIs and 27 did not have any prior experience. The data was recorded in a soundproof room with electromagnetic insulation. More information about the database can be found in \cite{dataset}. To reduce the ML models' training time, we analyzed only the frequencies of 12 Hz and 15 Hz. Although using only 2 classes may reduce the number of real-world applications of a BCI, it does not detract from the validity of the comparison of classifiers that we perform in this study. 

The second employed database is called BETA \cite{BETA}. We used it to perform cross-dataset testing, training ML models on Benchmark and testing them on BETA. We performed this procedure to better measure the generalization capability of our proposed methodologies. BETA is similar to Benchmark, it was also created with a virtual keyboard composed of 40 stimuli, with the same frequencies and phases observed in Benchmark. Furthermore, it also utilized 64-channel whole head EEG, with signals down-sampled from 1000 Hz to 250 Hz. A notch filter at 50Hz was employed again. Unlike Benchmark, the BETA dataset has 70 subjects, the first 15 observed 2 s stimuli, and the remaining ones, 3 s. Data acquisition was not performed in a soundproof or electromagnetically shielded room. The study had 4 blocks of 40 trials (one per stimulus frequency), and no subject was naive to BCIs. In this database we also only considered the 12 Hz and 15 Hz stimulation frequencies, allowing the cross-dataset testing.

Finally, we trained and tested the machine learning models in a third open database, which we will call Portable \ref{portable}. Its data was collected from a wearable SSVEP-based BCI, with 12 visual stimuli. They range from 9.25 Hz to 14.75 Hz (in 0.5 Hz increments), and adjacent stimuli have a phase difference of 0.5 $\pi$. We considered the data recorded by the wet Oz electrode, and the original database has signals from 8 electrodes. The original system sampling rate was 1000 Hz, but the signals were down-sampled to 250Hz by the dataset authors. The data was recorded in a room without electromagnetic shielding or sound insulation. There were 102 subjects in the study, who did not have previous experience with SSVEP-based BCIs. Each one participated in 10 blocks of 12 trials (one per target), and the trials had 2 s of stimulation time. A wearable interface can have worse quality and SNR in relation medical-grade stationary EEG systems. Therefore, with this test we expect to confirm if FBDNNs are also superior to alternative models when these limitations are considered. In the Portable dataset we will analyze the 12 target frequencies. Considering the utilization of a single-electrode and a small data length, the classification of 12 stimuli is an extreme test, which will not result in high accuracy. However, it will allow us to understand if FBDNNs are still advantageous (in relation to alternative models) when many targets are considered.

To simulate a single-channel BCI, we utilized only the data from the Oz electrode (according to the international 10-20 system). This electrode is placed above the visual cortex and, in a past study \cite{bassi_rampazzo_attux_2019}, we found that it produced better accuracies in relation to other electrodes (in that study we utilized a DNN and a SVM to analyze spectrograms of the EEG signals and obtained about 10\% better accuracy when utilizing only data from the Oz electrode, comparing to spectrograms created with signals from the electrodes O1, O2, Oz and POz). 

We are concerned with BCIs that do not require calibration in the final user (user-independent system). Therefore, when training the models, the test subject's data was not used in the training and validation datasets. When training on the Benchmark dataset, for each machine learning classification method, we trained one model for each of the 35 subjects considered as the test subject (leave-one-person-out cross-validation). The remaining data was randomly divided, with 75\% for training and 25\% for validation (hold-out). A similar approach was utilized for the Portable dataset, but we only considered its first 10 participants as test subjects, as creating models for each of the 102 subjects would be very computationally expensive. For the cross-dataset test, ML models were trained with the 35 subjects from Benchmark (still considering the random training and validation split), and then tested on all subjects in BETA.

\subsection{FBCCA}

Filter bank canonical correlation analysis (FBCCA) is a SSVEP classification method introduced in \cite{FBCCA}. CCA \cite{CCA} is a simple technique, commonly applied in SSVEP classification, due to its low computational cost, and high efficiency and robustness. The filter bank utilized in FBCCA allows CCA to take more advantage of the fundamental and harmonic components in SSVEP, and it is reported to have surpassed CCA accuracy \cite{FBCCA}.
 
Briefly, FBCCA starts by utilizing a filter bank with different band-pass filters to create multiple sub-band components of the EEG signals. For each stimulation frequency, we have a set of sinusoidal reference signals, whose frequencies are equal to the stimulation frequency and its harmonics. After filtering, we perform CCA between each created sub-band component and all sinusoidal reference signals. Finally, for each set of reference signals (corresponding to a visual stimulation frequency), we calculate a weighted sum of the squares of the correlation values corresponding to the sub-band components and the set. The weights in the sum are smaller for components containing higher frequency harmonics, because the SSVEP SNR is smaller for higher frequencies. The largest sum value indicates the SSVEP frequency \cite{FBCCA}.

\subsection{FBDNNs}
\label{MyNets}

The utilization of the filter bank can improve CCA extraction of stimulus-driven information from the harmonics and improve SSVEP classification performance \cite{FBCCA}. Our hypothesis in this study is that a filter bank would also allow different types of neural networks to more efficiently analyze the harmonic components of SSVEP and increase classification accuracy. 

Therefore, we suggest the utilization of FBDNNs, i.e., deep neural networks that have a filter bank in the signal processing pipeline. We tested this concept with different DNN architectures (FBCNN-2D, FBCNN-3D and the FBRNN), to understand if it can be regarded as a general preprocessing step for neural networks classifying SSVEP.

\subsubsection{Filter Bank}
Three different filter bank designs are presented in \cite{FBCCA}, M$_{1}$, M$_{2}$ and M$_{3}$. The M$_{1}$ method divides the frequency spectrum in consecutive sub-bands of the same width (10 sub-bands of 8 Hz, starting at 8 Hz). Unlike M$_{1}$, the M$_{2}$ method introduces superposition between the sub-bands, with the sub-band n starting at n x 8 Hz and ending at the minimum value between n x 16 Hz and 88 Hz.

We opted to use the filter bank design presented in the M$_{3}$ FBCCA method, in which sub-bands cover multiple harmonic frequency bands. The reason for this choice is that it produced the best results in \cite{FBCCA} and during our preliminary tests with the FBDNNs.

As in \cite{FBCCA}, all our filters are zero-phase Chebyshev type I infinite impulse response (IIR) filters. The passbands in the  M$_{3}$ method cover many harmonic frequency bands, and all of them have the same cutoff frequency, 90Hz, because \cite{FBCCA} analyzed that SSVEP harmonics had high SNR until around this frequency (with stimulation frequencies between 8 and 15.8 Hz). Table \ref{filterbank} shows the passbands of the filters in our filter bank. 

\begin{table}[!ht]
\centering
\begin{tabular}{|l|l|}
\hline
Filter & Passband          \\ \hline
1      & {[}6 Hz 90Hz{]}   \\ \hline
2      & {[}14 Hz 90 Hz{]} \\ \hline
3      & {[}22 Hz 90 Hz{]} \\ \hline
4      & {[}30 Hz 90 Hz{]} \\ \hline
5      & {[}38 Hz 90 Hz{]} \\ \hline
6      & {[}46 Hz 90 Hz{]} \\ \hline
7      & {[}54 Hz 90 Hz{]} \\ \hline
8      & {[}62 Hz 90 Hz{]} \\ \hline
9      & {[}70 Hz 90 Hz{]} \\ \hline
10     & {[}78 Hz 90 Hz{]} \\ \hline
\end{tabular}
\caption{Filters in the filter bank}
\label{filterbank}
\end{table}

Further tuning the filter bank configuration may improve results, but since in our preliminary tests changing the passbands and number of filters (e.g., using more filters and narrower passbands) only slightly changed the DNN accuracy, we decided to utilize the original M$_{3}$ filter bank. 

\subsubsection{FBCNN-2D}

For the FBCNN-2D, after creating the EEG sub-band components with the filter bank, we utilized fast Fourier transform (FFT). From the FFT result we removed frequencies above 90 Hz (since our cutoff frequency was 90 Hz). FFT (applied with the function scipy.fft.fftfreq) utilized a window length of 0.5 s and produced a frequency resolution of 2 Hz, resulting in 45 complex values between 0 and 90 Hz, whose real and imaginary parts we calculated.

The study in \cite{ComplexSSVEP} proposes a way to create the DNN input matrix from the FFT real and imaginary parts. They concatenate all the imaginary values after the real ones, creating a vector for each electrode. Afterwards, they stack these vectors to form their input matrix, \textbf{I}. This method is shown in equation 1, where X\textsubscript{O1}, X\textsubscript{Oz} and X\textsubscript{O2} are the EEG signals for the 3 electrodes used in \cite{ComplexSSVEP}.

\begin{equation}
\textbf{I}=	
\begin{bmatrix}
Re\{FFT(X\textsubscript{O1})\} & Im\{FFT(X\textsubscript{O1})\}\\
Re\{FFT(X\textsubscript{Oz})\} & Im\{FFT(X\textsubscript{Oz})\}\\
Re\{FFT(X\textsubscript{O2})\} & Im\{FFT(X\textsubscript{O2})\}
\end{bmatrix}
\end{equation}

With this matrix configuration, each complex number (related to each frequency interval in the spectrum) will have its real part in the first half of the matrix columns, and its imaginary part in the second. If the matrix has C columns, the real and imaginary parts of a complex number will be C/2 elements apart. This value can be fairly large, in our study, C=90 (and it would be even bigger with larger data lengths, which allow better frequency resolution in the FFT). Thus, only a very large convolutional kernel in the beginning of a DNN would be able to process the real and imaginary parts of a single complex number at a given position. Therefore, with this matrix configuration the first convolutional layers process the real and imaginary parts of the FFT separately.

We propose a different way of creating the input matrix, which consists of stacking the vectors representing the real and imaginary parts of the FFT result for each channel. With this method, the matrix in equation 1 would be reformulated in the manner depicted in equation 2.

\begin{equation}
\textbf{I}=	
\begin{bmatrix}
Re\{FFT(X\textsubscript{O1})\}\\
Im\{FFT(X\textsubscript{O1})\}\\
Re\{FFT(X\textsubscript{Oz})\}\\
Im\{FFT(X\textsubscript{Oz})\}\\
Re\{FFT(X\textsubscript{O2})\}\\
Im\{FFT(X\textsubscript{O2})\}
\end{bmatrix}
\end{equation}

Using this approach, the real and imaginary parts of each complex number given by the FFT are next to each other and small convolutional kernels are able to analyze both parts simultaneously. Therefore, we believe that this configuration is more adequate for a wider variety of CNN architectures and is more aligned with the concept of complex numbers. In the present work, we use only a single electrode (Oz), but we have 10 different sub-band components, created by the filter bank. Thus, instead of stacking the real and imaginary parts referent to different electrodes, we stack the parts that refer to the different sub-band components. We depicted the structure of the input matrix for the FBCNN-2D in equation 3, where X\textsubscript{SBn} represents the EEG signal $n$th sub-band component. Note that a row in matrix \textbf{I} (equation 3), $Re\{FFT(X\textsubscript{SBi})\}$ or $Im\{FFT(X\textsubscript{SBi})\}$, has multiple elements (in our case, 45). Thus, using 10 sub-band components, the matrix shall present 20 rows, and its shape shall be 20x45. The 45 size is due to our FFT operation, $FFT(X\textsubscript{SBi})$, which produces 45 values, representing frequencies from 0 to 90 Hz, with resolution of 2 Hz.

\begin{equation}
\textbf{I}=	
\begin{bmatrix}
Re\{FFT(X\textsubscript{SB1})\}\\
Im\{FFT(X\textsubscript{SB1})\}\\
Re\{FFT(X\textsubscript{SB2})\}\\
Im\{FFT(X\textsubscript{SB2})\}\\
...\\
Re\{FFT(X\textsubscript{SB10})\}\\
Im\{FFT(X\textsubscript{SB10})\}\\
\end{bmatrix}
\end{equation}

We normalize the matrix in (3), according to the dataset's standard deviation, computed over all sub-bands. It then serves as input for a two-dimensional convolutional neural network, the FBCNN-2D. In preliminary tests this input matrix produced higher accuracies with our DNN in relation to a configuration similar to the one in equation 1 (with distinct sub-band components in the different rows, instead of distinct electrodes).

The FBCNN-2D has 2 two-dimensional convolutional layers, each one followed by batch normalization, ReLU activation function, 2D max pooling (3x2 kernel after the first layer and 1x2 after the second) and dropout (25\%). At the end it has a fully-connected output layer, with softmax activation. 

\begin{figure}[!ht]
\caption{FBCNN-2D CNN structure.}
\includegraphics[width=1\textwidth]{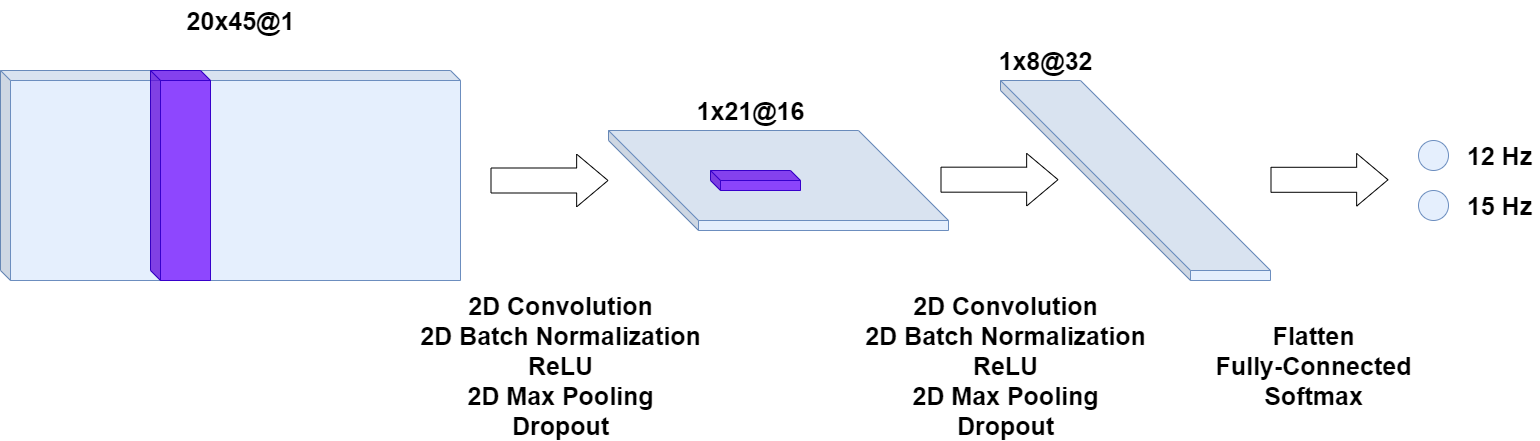}
\centering
\label{FigFBCNN2DS}
\end{figure}

The first convolutional layer has a kernel size of 20x6, with stride and padding of 1x1. Utilizing this kernel shape, the layer is able to ponder the contribution of every sub-band component, considering the real and imaginary parts of the complex numbers in each one of them. As a 2D convolution, it simultaneously performs frequency filtering. The layer has 16 channels. The input of the second convolution has the shape 1x21x16. It has 32 kernels of size 1x8 (with stride 1x1 and padding 0x1) and the layer processes the signal spectrum features. We used the sizes of 8 and 6 in the convolutional kernel shapes to provide a relatively large receptive field in the frequency dimension, after the convolutional and max pooling layers. We used batch normalization and dropout to avoid overfitting and improve generalization, having in mind that our BCI is user-independent. Batch normalization was conceived to reduce the problem of internal covariance shift, i.e., the change in the distribution of DNN layer inputs during training, due to parameter updates (which makes gradient descent more difficult)\cite{ioffe2015batch}. However, it also makes the DNN output for a single example non-deterministic, adding a regularization effect \cite{ioffe2015batch}. Furthermore, a recent work has shown that it can improve CNN accuracies with SSVEP classification \cite{BnSSVEP}. The FBCNN-2D ends with a fully-connected layer containing 2 or 12 neurons (for Benchmark/Beta or Portable datasets) and softmax activation, in order to produce probability estimations for the classes. We employed preliminary tests and evaluation of validation accuracy to fine tune hyper-parameters, such as dropout, the exact kernel size and number of convolutional channels. The DNN is represented in figure \ref{FigFBCNN2DS}, bright shapes represent input, feature maps and outputs, while the darker cuboids represent convolutional kernel sizes. The arrows indicate the DNN operations and layers. Numbers above the cuboids are in the following format: height x length @ channels.

\begin{figure}[!ht]
\caption{FBCNN-2D signal processing.}
\includegraphics[width=1\textwidth]{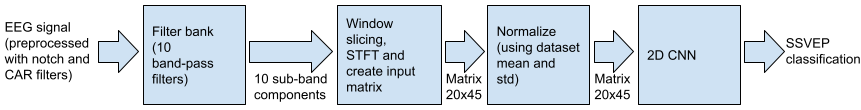}
\centering
\label{FigFBCNN2D}
\end{figure}

Figure \ref{FigFBCNN2D} shows the entire FBCNN-2D pipeline, including the filter bank and all signal preprocessing stages. Before the bank, we have a notch filter at 50 Hz in the stationary EEG datasets (to remove power line noise, applied by the datasets' authors \cite{dataset} \cite{BETA}), followed by a common average reference (CAR) filter (for the Benchmark and BETA datasets) or a band-pass filter between 2 Hz and 90 Hz (for the portable dataset). We utilized the CAR/band-pass and notch filters with all the classification methods in this study (unlike our previous work \cite{BSPCWork}, which did not use the CAR/band-pass filter with FBCCA). Window slicing (WS) is applied after the filter bank, it segments the signals in the time dimension, with superposition, and it serves as a form of data augmentation and to determine the BCI data length. WS used a window of size 0.5 s (simulating a high-speed and low latency BCI and allowing enough frequency resolution to distinguish between the 12 Hz and 15 Hz frequencies), with displacement of 0.1 s (a small value, to augment the dataset).  We note that, if we apply WS before the filter bank, we will filter very short signals (0.5 s in this study), degrading them and reducing the BCI performance. Therefore, in an online BCI, the filter bank should filter the data stream (like the notch filter), before segmenting the signal to analyze it with the DNN. Every classification method in this study used window slicing with the same parameters, so they could analyze signals with the same data length (0.5 s). 

\subsection{FBCNN-3D}

The inputs of the FBCNN-3D are complex spectrograms. To create them, we applied a short-time Fourier transform (STFT) to each sub-band component of the EEG signal. STFT (using the function scipy.signal.signal.stft) utilized a rectangular window (improving frequency resolution) of length 125 (0.5 s) and hop length of 62 (0.248 s). Frequency resolution after STFT was 2 Hz and time resolution 0.167 s. Thus, each STFT operation produced a complex matrix of size 45x3 (representing frequencies from 0 to 90 Hz and time from 0 to 0.5 s). There is one matrix for each sub-band component and the DNN input is a 3D tensor composed of these matrices. Following the reasoning presented in the previous section, about the FBCNN-2D, we want the real and imaginary parts of each complex number produced by the STFT to be next to each other. To accomplish this, we could alternate between real and imaginary parts in any of the three dimensions of the input tensor: the time dimension, the frequency dimension or the sub-band dimension. With preliminary tests and analysis of DNN validation error, we observed the best results alternating the values in the time dimension. 

We represent the output of a STFT operation on the n-th sub-band component of the EEG signal ($X_{SBn}$) as a matrix, $\textbf{STFT}^{SBn}$, with elements $c_{i,j}^{SBn}$, which are complex numbers. In this study the matrix has shape 45x3, where 45 refers to the frequency dimension (selected between 0 and 90 Hz, with a resolution of 2 Hz) and 3 to the time dimension (between 0 and 0.5 s, with resolution of 0.167 s). 

\begin{equation}
STFT(X_{SBn})=\textbf{STFT}^{SBn}=
\begin{bmatrix}
c_{1,1}^{SBn} & c_{1,2}^{SBn} & c_{1,3}^{SBn}\\
c_{2,1}^{SBn} & c_{2,2}^{SBn} & c_{2,3}^{SBn}\\
... & ... & ...\\
c_{45,1}^{SBn} & c_{45,2}^{SBn} & c_{45,3}^{SBn}\\
\end{bmatrix}
\end{equation}

Alternating between real and imaginary values in the time dimension, our input matrix for the n-th sub-band, $\textbf{I}^{SBn}$, has the format shown in equation 5:

\begin{equation}
\textbf{I}^{SBn}=
\begin{bmatrix}
Re\{c_{1,1}^{SBn}\} & Im\{c_{1,1}^{SBn}\} & Re\{c_{1,2}^{SBn}\} & Im\{c_{1,2}^{SBn}\} & Re\{c_{1,3}^{SBn}\} & Im\{c_{1,3}^{SBn}\} \\
Re\{c_{2,1}^{SBn}\} & Im\{c_{2,1}^{SBn}\} & Re\{c_{2,2}^{SBn}\} & Im\{c_{2,2}^{SBn}\} & Re\{c_{2,3}^{SBn}\} & Im\{c_{2,3}^{SBn}\} \\
... & ... & ... & ... & ... & ...\\
Re\{c_{45,1}^{SBn}\} & Im\{c_{45,1}^{SBn}\} & Re\{c_{45,2}^{SBn}\} & Im\{c_{45,2}^{SBn}\} & Re\{c_{45,3}^{SBn}\} & Im\{c_{45,3}^{SBn}\} \\\\
\end{bmatrix}
\end{equation}

We stack the 10 $\textbf{I}^{SBn}$ matrices, referent to the 10 sub-band components of the EEG signal, creating the FBCNN-3D input tensor, $\textbf{I}$, with shape 10x45x6 (whose dimensions represent sub-band component, frequency and time, respectively). This tensor is then normalized.

The tensor $\textbf{I}$ is the input for a 3D convolutional DNN. The FBCNN-3D has two three-dimensional convolutional layers, followed by a fully connected layer. Each convolutional layer is followed by 3D batch normalization, leaky ReLU activation function, 3D max pooling (2x2x3 kernel after the first layer and 3x2x1 after the second) and dropout (25\% probability). Again, the DNN ends with a fully-connected layer, with softmax activation and 2 or 12 artificial neurons (depending on the dataset being analyzed). 

\begin{figure}[!ht]
\caption{FBCNN-3D CNN structure.}
\includegraphics[width=1\textwidth]{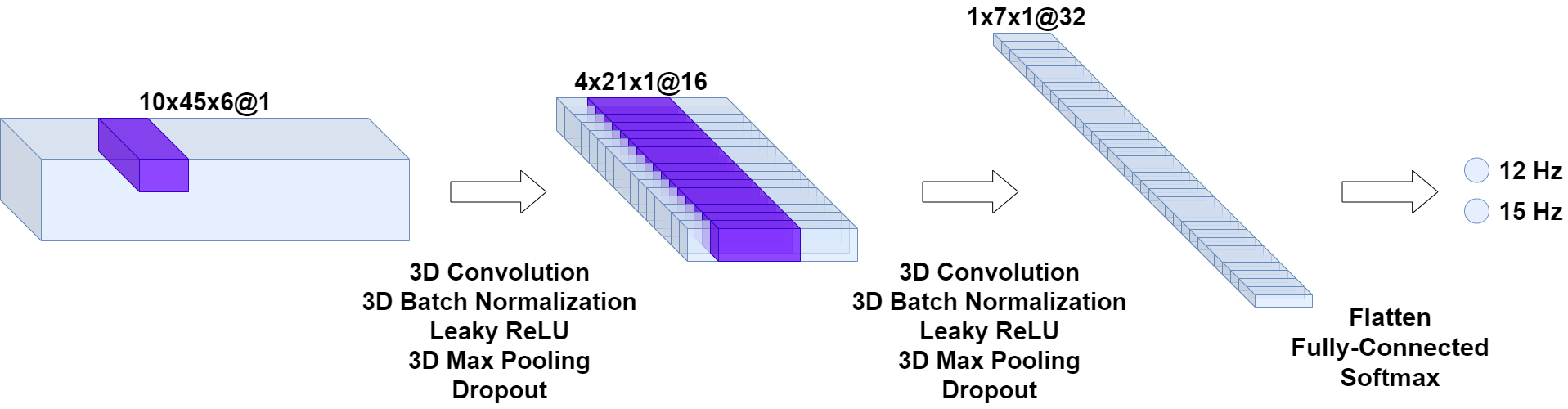}
\centering
\label{FigFBCNN3DS}
\end{figure}

The first convolutional layer has 16 kernels of shape 4x6x6 (with stride and padding 1x1x1). Thus, it performs a 3D convolution that operates simultaneously on the three dimensions (sub-bands, frequency and time). We observed superior performances with this configuration, in relation to convolutions that did not process all dimensions simultaneously. We note that this convolution filter can already cover the entire time dimension (of size 6), in which we alternated the real and imaginary parts of the STFT complex numbers. This approach is similar to the FBCNN-2D's, where we alternated real and imaginary parts in the sub-band dimension and used a convolutional kernel in the first layer that covered it entirely. The second 3D convolutional layer receives an input of shape 4x21x1x16 (16 channels) and has 32 kernels of size 4x10x1 (with stride 1x1x1 and padding 1x1x0). We chose this format to reduce the feature map first and third dimensions (sub-bands and time) to 1 element (after max pooling) and to generate a large receptive field in the second dimension (frequency). Exact kernel shapes, numbers of channels, dropout percentage and other hyper-parameters were chosen after observing validation accuracy in preliminary tests. As in the 2D CNN, we applied batch normalization and dropout to avoid overfitting and improve generalization. We represent the network structure in figure \ref{FigFBCNN3DS}. Again, bright structures represent input, feature maps or output, and dark shades represent kernel size. Numbers above the cuboids are in the following format: height x length x depth @ channels. Notice that, considering the channels, we have 4-dimensional tensors. In the figure, the first 3 dimensions are represented by the dimensions of a single cuboid, while the channel dimension is represented by the adjacent copies of a cuboid.

\begin{figure}[!ht]
\caption{FBCNN-3D signal processing.}
\includegraphics[width=1\textwidth]{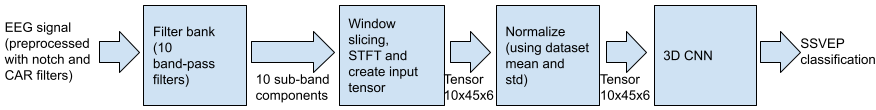}
\centering
\label{FigFBCNN3D}
\end{figure}

Figure \ref{FigFBCNN3D} shows how the FBCNN-3D works, we note that the differences between it and the FBCNN-2D pipeline are the utilization of STFT instead of FFT, a tensor instead of a matrix as the DNN input and the 3D CNN instead of the 2D network. The other preprocessing stages (notch and CAR/band-pass filter, filter bank, WS and normalization) were conducted in the same way for both DNNs, to allow a better comparison of the two networks.

\subsubsection{FBRNN}

Unlike the CNNs, the FBRNN, a deep recurrent neural network composed of 1D convolutional layers followed by LSTM layers, analyzes the sub-band components of the EEG signal directly, in the time domain. Therefore, its inputs are 10 vectors ($X_{SB1}$,...,$X_{SB10}$) representing the 10 sub-band components, which the RNN sees as a one-dimensional input with 10 channels. The signals have a duration of 0.5 s, therefore, with the sampling rate of 250Hz, the vectors' length is 125. Before feeding them to the neural network, we normalize the vectors.

The RNN structure combines 1D convolutions, which act as time filters, and LSTM layers, recurrent structures that are able to capture short and long-term dependencies in the time domain \cite{LSTM}. We performed preliminary tests with different DNN depths, and obtained the best validation results with a network composed of two convolutional layers, followed by five LSTM layers and finishing with a fully-connected layer (composed of 2 or 12 artificial neurons). The output layer accesses the last LSTM layer output features for every time step, because only the last response is not enough to efficiently classify the SSVEP signal. Both convolutional layers are followed by batch normalization (1D), a ReLU activation function, max pooling (1D, with kernel size of 2) and dropout (40\% probability). In this network dropout and batch normalization are also employed to prevent overfitting.

\begin{figure}[!ht]
\caption{FBRNN DNN structure.}
\includegraphics[width=1\textwidth]{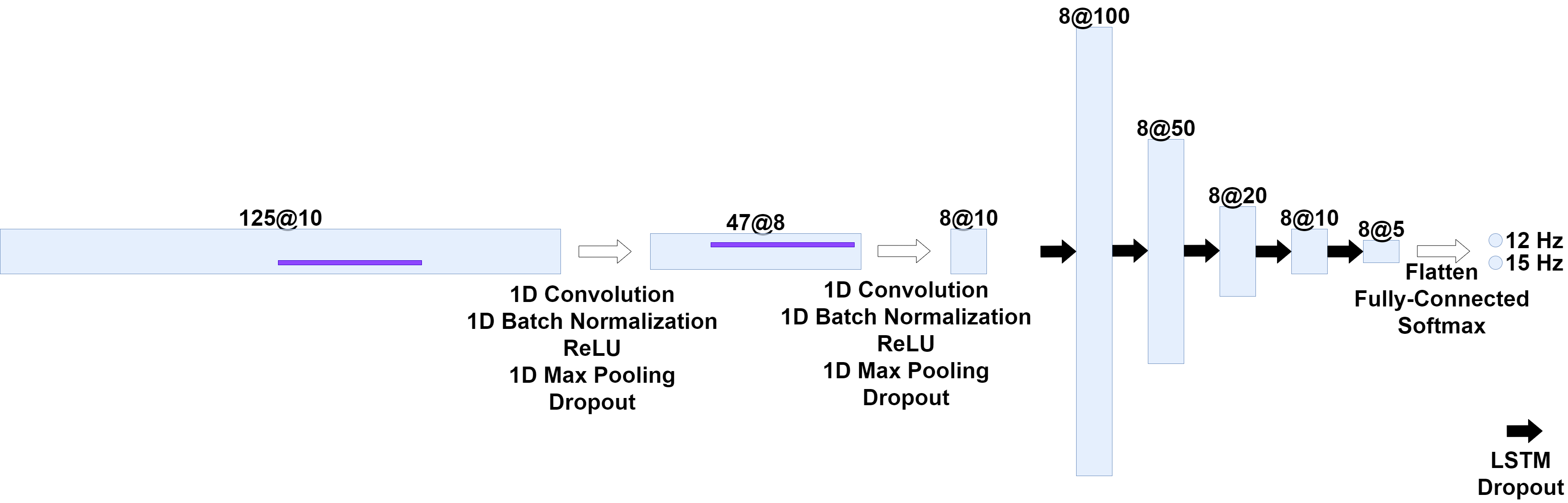}
\centering
\label{FigFBRNNS}
\end{figure}

Both convolutional layers have kernel size of 32, which generates a large receptive field and allows the convolutional structure to capture long-term dependencies. Indeed, our preliminary tests showed that smaller kernels reduced validation accuracy. The first layer has 8 channels, and the second, 10. Therefore, there is little change in the number of channels, which is 10 in the input. The convolutional layers are followed by 5 LSTM layers, in which the number of hidden features decrease by each consecutive layer (the 5 consecutive layers have 100, 50, 20, 10 and 5 hidden features, respectively). Each LSTM layer is followed by dropout (40\%). Finally, the output layer contains two neurons, with softmax activation, to predict the class probabilities. Exact kernel shapes and number of channels, dropout and number of LSTM hidden features were selected with the help of preliminary tests and evaluation of validation error. Figure \ref{FigFBRNNS} depicts the DNN structure, black arrows indicate the LSTM layers. Again, bright shapes represent the layers' inputs and outputs and dark rectangles show convolutional kernel sizes. Numbers above the rectangles are in the following format: length (time sequence) @ channels/hidden features.

\begin{figure}[!ht]
\caption{FBRNN architecture and signal processing.}
\includegraphics[width=1\textwidth]{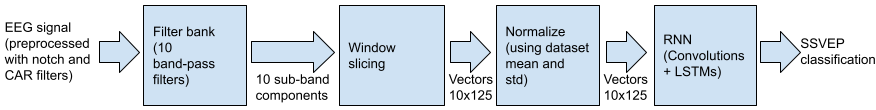}
\centering
\label{FigFBRNN}
\end{figure}

Figure \ref{FigFBRNN} displays all the FBRNN signal processing steps. We applied the same preprocessing steps that were used with the FBCNNs (notch and CAR/band-pass filters, filter bank, window slicing and normalization), the only difference is not utilizing FFT or STFT. 

\subsection{Alternative classification methods}

In this study, besides the FBDNNs, we also employed other SSVEP classification methods to better compare our proposed neural networks to the most often employed models and the state-of-the-art.

The first alternative SSVEP classification method that we used is a support vector machine. These models are a very popular method for SSVEP classification (e.g. \cite{SVMEx} and \cite{SVMExComparison}) and they are shallow and relatively easy to train. We chose a linear SVM kernel because it is commonly used and because other works \cite{SVMExComparison} did not find a significant improvement with alternative kernels (their best results were achieved with RBF kernels, only increasing accuracy by 0.13\% in relation to the linear one). Our SVM analyzed the Oz electrode data in the frequency domain (utilizing FFT). The analysis of complex spectrum features is not usual and, as far as we know, is not used with SVMs in the field of SSVEP classification. Therefore, in order to compare the FBDNNs to commonly used techniques, our SVM analyzed the magnitude spectrum. Thus, the Oz electrode signal was preprocessed with the notch (for stationary EEG datasets) and CAR/band-pass filter, window slicing was applied and followed by FFT. The parameters of these preprocessing steps are the same as in the FBDNNs. To create the SVM input vector the magnitude of the FFT output was transferred to the Decibels scale and normalized. We trained a Random Forest (RF) with the same preprocessing pipeline (and inputs) employed for the SVM. The classifier had 100 trees, and in the Benchmark and Portable datasets, 11 and 14 features were considered for node splits, respectively. These values were chosen according to validation error, and the other parameters were set as the Scikit Learn library defaults (sklearn.ensemble.RandomForestClassifier).

We also classified the dataset with FBCCA, the technique that inspired the FBDNNs and surpassed CCA \cite{FBCCA}. We performed FBCCA with the same parameters utilized to classify the original Benchmark dataset in \cite{dataset}: 7 sub-band components, analyzing 5 harmonics, a=1.25 and b=0.25 (a and b determine the weights in FBCCA weighted sum of squares of the correlation values \cite{FBCCA}). We created the 7 sub-band components using the first seven filters in table \ref{filterbank}. FBCCA obtained superior results with 7 filters, whereas the FBDNNs showed slightly superior performances with 10, according to our preliminary tests. For this method, the EEG signals were also preprocessed with the notch and CAR/band-pass filters and window slicing, employing again the same parameters used for the FBDNNs.

To better understand the benefits of the filter bank, we trained a CNN without it. This DNN, which we will call A-CNN (alternative CNN), was trained to analyze complex spectrograms of the Oz electrode signal. We created these objects in the way depicted in equations 4 and 5, but considering the signal ($X$) instead of its multiple sub-band components ($X_{SBn}$). Therefore, equations 6 and 7 represent the creation of the A-CNN input, \textbf{I}.

\begin{equation}
STFT(X)=\textbf{STFT}=
\begin{bmatrix}
c_{1,1} & c_{1,2} & c_{1,3}\\
c_{2,1} & c_{2,2} & c_{2,3}\\
... & ... & ...\\
c_{45,1} & c_{45,2} & c_{45,3}\\
\end{bmatrix}
\end{equation}

\begin{equation}
\textbf{I}=
\begin{bmatrix}
Re\{c_{1,1}\} & Im\{c_{1,1}\} & Re\{c_{1,2}\} & Im\{c_{1,2}\} & Re\{c_{1,3}\} & Im\{c_{1,3}\} \\
Re\{c_{2,1}\} & Im\{c_{2,1}\} & Re\{c_{2,2}\} & Im\{c_{2,2}\} & Re\{c_{2,3}\} & Im\{c_{2,3}\} \\
... & ... & ... & ... & ... & ...\\
Re\{c_{45,1}\} & Im\{c_{45,1}\} & Re\{c_{45,2}\} & Im\{c_{45,2}\} & Re\{c_{45,3}\} & Im\{c_{45,3}\} \\\\
\end{bmatrix}
\end{equation}

The A-CNN has the same network structure used in the FBCNN-3D (depicted in figure \ref{FigFBCNN3DS}), excluding the dimension associated with the sub-band components (the first dimension in the kernels, feature maps, inputs, padding and strides). Therefore, this network has two 2D convolutional layers, with kernel shapes of 6x6 and 10x1. Each one of them is followed by batch normalization (2D), ReLU activation, max pooling (2D) and dropout (25\%). Its output layer is fully-connected, with 2 or 12 (according to the dataset) neurons and softmax activation. We chose this architecture to provide a better comparison with the FBCNN-3D, considering that both analyze complex spectrograms, but one utilizes a filter bank. All other signal preprocessing stages for the FBCNN-3D and the A-CNN are the same (car/band-pass and notch filters, window slicing, STFT and normalization, as shown in figure \ref{FigFBCNN3D}).

We also trained a RNN without the filter bank, which we will call A-RNN. The network receives a single vector as input (the EEG signal, $X$, in the time domain), and has the same structure of the FBRNN (figure \ref{FigFBRNNS}), with the only difference being the first layer's input shape (125 elements and 10 channels in the FBRNN and 125 elements with a single channel in the A-RNN). Both networks have the same signal preprocessing stages (as figure \ref{FigFBRNN} shows, except for the filter bank).

Therefore, we trained a DNN that is very similar to the FBCNN-3D (the A-CNN) and another that is almost identical to the FBRNN (the A-RNN). In this way, we aim to better understand the contributions of filter banks to convolutional neural networks and to recurrent NNs. 

\subsection{Training Procedure}
%svm, 5 dnns
We implemented all our models in Python, utilizing the PyTorch library to train the neural networks. The computer used in this study had a NVidia RTX 3080 graphics power unit.

We utilized the same training procedures with all the CNNs (the A-CNN, FBCNN-2D and FBCNN-3D), to better compare them. We trained the models with stochastic gradient descent (SGD), using momentum of 0.9, cross-entropy loss and learning rate of 0.001. A maximum of 1000 epochs was defined for the Benchmark dataset, and 2000 for Portable. But it was never achieved, due to early-stopping. We used two stopping criteria, 50/250 epochs (for Benchmark/Portable) without improving validation loss (patience of 50 or 250) or 15 epochs, if the training loss was low (below 0.1) and the validation loss very high (0.25 above the lowest validation loss obtained in the previous epochs). The second criterion reduced training time when the DNN had already achieved a strong overfitting. We employed mini-batches of 16 samples in the Benchmark dataset and 64 in the Portable database.

Due to their different nature, we trained the RNNs with different parameters in relation to the CNNs, but we trained the A-RNN and the FBRNN in the same manner. We utilized the Adam optimizer, with betas of 0.9 and 0.999, and epsilon of 10$^{-8}$. We employed a learning rate of 0.001, cross-entropy loss, mini-batches of 16 or 64 (Benchmark/Portable) samples, a maximum of 10000 epochs and early stopping with patience of 300.

Finally, for the SVM we used SGD with momentum of 0.9, learning rate of 0.001 and mini-batches of 16 or 64 (Benchmark/Portable) samples. We chose a maximum of 3000 epochs (also not achieved), and we used early-stopping, with patience of 200. We trained with the hinge loss and the regularization parameter c=0. For the Portable dataset, the hinge loss was replaced by the multi-class hinge loss, to consider the 12 targets.

\section{Results}

In the Benchmark dataset \cite{dataset}, following the user-independent approach, we created a classification model for each BCI user as the test subject (leave-one-person-out cross-validation). In table \ref{res} we present the test accuracies for each one of them, along with the mean accuracy and mean F1-Score, considering all the 35 test subjects. Ranking the classification models according to best mean accuracy or mean F1-Score, we obtain the following order: FBRNN, FBCNN-3D, FBCNN-2D, A-CNN, A-RNN, SVM and FBCCA. We note that the the high standard deviations (std) observed in table \ref{res} are expected, due to the SSVEP's intrinsic high variability between subjects \cite{variability}.

\begin{table}[]
\centering
\begin{tabular}{|l|l|l|l|l|l|l|l|l|}
\hline
Subjects                                                         & FBRNN & FBCNN-3D & FBCNN-2D & A-CNN & A-RNN & SVM   & FBCCA & \begin{tabular}[c]{@{}l@{}}Random\\ Forest\end{tabular} \\ \hline
1                                                                & 83.7  & 82.8     & 81.9     & 77.7  & 78.1  & 80.8  & 86.8  & 54.3                                                    \\ \hline
2                                                                & 87.9  & 87.7     & 87       & 83.7  & 79.5  & 82.6  & 87.9  & 55.3                                                    \\ \hline
3                                                                & 97.6  & 97.3     & 97.5     & 95.3  & 94.9  & 93.7  & 94.7  & 46.6                                                    \\ \hline
4                                                                & 78.4  & 77       & 78.1     & 75.4  & 74.8  & 69.6  & 70.3  & 57.8                                                    \\ \hline
5                                                                & 97.6  & 96.2     & 96.2     & 93.8  & 93.5  & 89.7  & 96.6  & 55.1                                                    \\ \hline
6                                                                & 96    & 95.5     & 96.7     & 93.5  & 93.8  & 90.6  & 87    & 49.5                                                    \\ \hline
7                                                                & 91.8  & 93.1     & 92.4     & 88.8  & 88    & 86.6  & 88.4  & 55.8                                                    \\ \hline
8                                                                & 96.7  & 96.6     & 96.4     & 93.7  & 95.3  & 91.3  & 88.8  & 56.3                                                    \\ \hline
9                                                                & 86.8  & 81.9     & 81.2     & 82.8  & 78.6  & 79.2  & 80.4  & 48.9                                                    \\ \hline
10                                                               & 92.8  & 92       & 94.4     & 90.4  & 89.9  & 86.4  & 90.8  & 55.3                                                    \\ \hline
11                                                               & 70.5  & 67.9     & 66.3     & 65.6  & 65.8  & 62.1  & 56.2  & 50.4                                                    \\ \hline
12                                                               & 78.3  & 77.7     & 78.8     & 63.8  & 73.4  & 66.7  & 62.9  & 48.6                                                    \\ \hline
13                                                               & 61.8  & 59.8     & 59.8     & 59.1  & 57.8  & 58    & 56    & 48.6                                                    \\ \hline
14                                                               & 76.3  & 77.9     & 76.4     & 70.5  & 70.8  & 67.6  & 65.9  & 66.8                                                    \\ \hline
15                                                               & 92.2  & 94.6     & 94.6     & 90    & 88.9  & 88.4  & 84.1  & 55.6                                                    \\ \hline
16                                                               & 87    & 85.5     & 80.4     & 79    & 75.7  & 72.3  & 79    & 52.2                                                    \\ \hline
17                                                               & 80.3  & 78.3     & 79.9     & 76.6  & 78.1  & 70.5  & 73.6  & 50.2                                                    \\ \hline
18                                                               & 83.5  & 78.6     & 79.2     & 78.6  & 76.8  & 77.2  & 77.4  & 51.6                                                    \\ \hline
19                                                               & 80.8  & 82.1     & 78.4     & 75.9  & 73    & 75.4  & 70.3  & 52.7                                                    \\ \hline
20                                                               & 93.7  & 92.6     & 94.4     & 92.2  & 92    & 86.8  & 90.2  & 53.3                                                    \\ \hline
21                                                               & 78.6  & 79.5     & 77.2     & 75.5  & 75    & 75.2  & 73.7  & 58.7                                                    \\ \hline
22                                                               & 98.7  & 97.6     & 96.6     & 95.8  & 95.7  & 94.2  & 83.5  & 49.8                                                    \\ \hline
23                                                               & 77    & 75.5     & 74.8     & 71.6  & 69.4  & 65.2  & 71.6  & 52.2                                                    \\ \hline
24                                                               & 97.1  & 96.9     & 97.3     & 95.3  & 92.2  & 91.5  & 91.8  & 57.1                                                    \\ \hline
25                                                               & 92.4  & 94.2     & 93.7     & 84.6  & 86.2  & 81.5  & 76.8  & 57.6                                                    \\ \hline
26                                                               & 93.1  & 94.7     & 94.2     & 91.7  & 90.2  & 89.1  & 77.2  & 65.6                                                    \\ \hline
27                                                               & 97.1  & 95.5     & 96       & 93.3  & 89.9  & 89.5  & 95.1  & 50.2                                                    \\ \hline
28                                                               & 96.2  & 95.5     & 95.5     & 94.4  & 92.4  & 90.8  & 96.4  & 56.3                                                    \\ \hline
29                                                               & 73.4  & 72.6     & 71.2     & 54.3  & 60    & 58.3  & 53.3  & 57.2                                                    \\ \hline
30                                                               & 97.3  & 96.9     & 96.4     & 94.4  & 95.1  & 93.1  & 93.5  & 53.4                                                    \\ \hline
31                                                               & 94.4  & 94.6     & 96.2     & 93.5  & 93.8  & 90.8  & 86.1  & 58                                                      \\ \hline
32                                                               & 99.1  & 98.9     & 99.5     & 98.7  & 98.6  & 96.6  & 97.1  & 60                                                      \\ \hline
33                                                               & 65.4  & 68.5     & 66.8     & 69.6  & 67.4  & 72.3  & 71.4  & 56.2                                                    \\ \hline
34                                                               & 86.2  & 84.2     & 84.6     & 78.8  & 78.6  & 77.7  & 74.3  & 50.5                                                    \\ \hline
35                                                               & 95.7  & 93.7     & 93.1     & 90.8  & 92.9  & 89.1  & 77.4  & 62.7                                                    \\ \hline
\textbf{\begin{tabular}[c]{@{}l@{}}Mean\\ Accuracy\end{tabular}} & 87.3  & 86.7     & 86.4     & 83.1  & 82.7  & 80.9  & 80.2  & 54.6                                                    \\ \hline
\textbf{\begin{tabular}[c]{@{}l@{}}Accuracy\\ Std\end{tabular}}  & 10    & 10.2     & 10.7     & 11.6  & 11.1  & 10.9  & 12    & 4.65                                                    \\ \hline
\textbf{\begin{tabular}[c]{@{}l@{}}Mean\\ F1-Score\end{tabular}} & 0.877 & 0.872    & 0.867    & 0.831 & 0.829 & 0.807 & 0.786 & 0.544                                                   \\ \hline
\textbf{\begin{tabular}[c]{@{}l@{}}F1-Score\\ Std\end{tabular}}  & 0.093 & 0.097    & 0.106    & 0.115 & 0.11  & 0.113 & 0.135 & 0.088                                                   \\ \hline
\end{tabular}
\caption{Test accuracies (\%) for test subjects, mean accuracies (\%) and mean F1-Scores, in the Benchmark dataset.}
\label{res}
\end{table}

To ensure that our approach has good generalization capacity, we tested the models, trained on the Benchmark dataset, on an independent database. The chosen dataset is the BETA \cite{BETA}, with 70 subjects. We tested the models on each subject, and calculated the mean performances, reported in table \ref{BetaTest}. As expected when training models on one dataset and testing on another, all trainable models had an accuracy reduction. The FBDNNs still performed better than the alternative models. FBCCA surpassed the neural networks without filter banks, as the technique does not employ training.

\begin{table}[]
\centering
\begin{tabular}{|l|l|l|l|l|l|l|l|l|}
\hline
                                                                         & FBRNN & FBCNN-3D & FBCNN-2D & A-CNN & A-RNN & SVM   & FBCCA & \begin{tabular}[c]{@{}l@{}}Random\\ Forest\end{tabular} \\ \hline
\begin{tabular}[c]{@{}l@{}}Mean\\ Accuracy\end{tabular}                  & 79.9  & 79.5     & 78.9     & 74.6  & 69.7  & 73.7  & 77.5  & 52.5                                                    \\ \hline
\begin{tabular}[c]{@{}l@{}}Accuracy \\ Std\end{tabular} & 13.8  & 13       & 13.7     & 13.6  & 13.3  & 13.1  & 13.8  & 5.7                                                     \\ \hline
\begin{tabular}[c]{@{}l@{}}Mean\\ F1-Score\end{tabular}                  & 0.794 & 0.802    & 0.791    & 0.729 & 0.703 & 0.718 & 0.767 & 0.506                                                   \\ \hline
\begin{tabular}[c]{@{}l@{}}F1-Score\\ Std\end{tabular}  & 0.154 & 0.128    & 0.148    & 0.156 & 0.135 & 0.154 & 0.153 & 0.097                                                   \\ \hline
\end{tabular}
\caption{Mean test accuracies (\%) and F1-Scores in the BETA dataset, for models trained in the Benchmark dataset.}
\label{BetaTest}
\end{table}

Finally, as an extreme test, we utilized the machine learning models to classify the Portable SSVEP dataset \cite{portable}. The results are shown in table \ref{portable}. Accuracy is much lower, because we are using a single electrode with a small data length (0.5 s) to classify 12 targets in a wearable interface. However, we can see that, for the third time, the FBDNNs surpassed the common DNNs. Furthermore, all neural networks performances were better than the other machine learning models. Relative to the previous experiments, the RNNs saw a performance reduction, being surpassed by the CNNs. Thus, in the portable dataset, the best performing model was the FBCNN-3D.

\begin{table}[!h]
\centering
\begin{tabular}{|l|l|l|l|l|l|l|l|l|}
\hline
Subjects                                                         & FBRNN & FBCNN-3D & FBCNN-2D & A-CNN & A-RNN & SVM   & FBCCA & \begin{tabular}[c]{@{}l@{}}Random\\ Forest\end{tabular} \\ \hline
1                                                                & 13.3  & 19.9     & 17.9     & 15.8  & 14.1  & 13.9  & 10.1  & 12                                                      \\ \hline
2                                                                & 11.2  & 12.7     & 12.6     & 12.2  & 10.4  & 10.6  & 8.13  & 9.9                                                     \\ \hline
3                                                                & 21.2  & 27.3     & 25       & 21.3  & 16.7  & 15.6  & 9.64  & 14.8                                                    \\ \hline
4                                                                & 25.3  & 42.2     & 34.7     & 29.5  & 20.3  & 17.7  & 9.84  & 16.6                                                    \\ \hline
5                                                                & 15.2  & 21.8     & 21.4     & 18.5  & 14.8  & 15.4  & 10.2  & 15                                                      \\ \hline
6                                                                & 11    & 12.1     & 12.6     & 10.9  & 10.3  & 9.2   & 8.85  & 10.2                                                    \\ \hline
7                                                                & 16.6  & 22.8     & 20.3     & 18.6  & 17    & 14.1  & 11.5  & 12.6                                                    \\ \hline
8                                                                & 18.2  & 20.9     & 19.5     & 15.9  & 15.4  & 11.8  & 10.8  & 9.7                                                     \\ \hline
9                                                                & 17.7  & 20.6     & 20.2     & 15.7  & 14.4  & 10.5  & 11.8  & 11.3                                                    \\ \hline
10                                                               & 16    & 14.6     & 15.9     & 14.5  & 15.2  & 10.5  & 8.49  & 9.5                                                     \\ \hline
\textbf{\begin{tabular}[c]{@{}l@{}}Mean\\ Accuracy\end{tabular}} & 16.6  & 21.5     & 20       & 17.3  & 14.8  & 12.9  & 9.93  & 12.2                                                    \\ \hline
\textbf{\begin{tabular}[c]{@{}l@{}}Accuracy\\ Std\end{tabular}}  & 4.18  & 8.26     & 6.12     & 4.98  & 2.81  & 2.67  & 1.15  & 2.39                                                    \\ \hline
\textbf{\begin{tabular}[c]{@{}l@{}}Mean\\ F1-Score\end{tabular}} & 0.14  & 0.207    & 0.194    & 0.165 & 0.125 & 0.107 & 0.056 & 0.117                                                   \\ \hline
\textbf{\begin{tabular}[c]{@{}l@{}}F1-Score\\ Std\end{tabular}}  & 0.052 & 0.084    & 0.068    & 0.053 & 0.033 & 0.029 & 0.033 & 0.025                                                   \\ \hline
\end{tabular}
\caption{Test accuracies (\%) for test subjects, mean accuracies (\%) and mean F1-Scores, in the Portable dataset.}
\label{portable}
\end{table}

Table \ref{statistics} shows statistical tests comparing pairs of different models. Each cell row and column indicate the two methods whose accuracies are being compared. We employed paired Wilcoxon signed-rank tests, and the cells display the resulting p-values. The table shows comparisons among the neural networks, and compares them to the remaining methodologies (SVM, FBCCA and Random Forest).

\begin{table}[h]
\centering
\begin{tabular}{|llllll|}
\hline

\multicolumn{1}{|l|}{-}             & \multicolumn{1}{l|}{FBRNN}       & \multicolumn{1}{l|}{FBCNN-3D}    & \multicolumn{1}{l|}{FBCNN-2D}    & \multicolumn{1}{l|}{A-RNN}       & A-CNN       \\ \hline
\multicolumn{6}{|c|}{\textbf{Benchmark Dataset}}                                                                                                                                              \\ \hline
\multicolumn{1}{|l|}{FBRNN}         & \multicolumn{1}{l|}{-}           & \multicolumn{1}{l|}{4.14e-02*}   & \multicolumn{1}{l|}{2.43e-02*}   & \multicolumn{1}{l|}{3.81e-07***} & 1.59e-06*** \\ \hline
\multicolumn{1}{|l|}{FBCNN-3D}      & \multicolumn{1}{l|}{4.14e-02*}   & \multicolumn{1}{l|}{-}           & \multicolumn{1}{l|}{3.08e-01}    & \multicolumn{1}{l|}{2.47e-07***} & 8.84e-07*** \\ \hline
\multicolumn{1}{|l|}{FBCNN-2D}      & \multicolumn{1}{l|}{2.43e-02*}   & \multicolumn{1}{l|}{3.08e-01}    & \multicolumn{1}{l|}{-}           & \multicolumn{1}{l|}{3.21e-07***} & 3.03e-06*** \\ \hline
\multicolumn{1}{|l|}{A-RNN}         & \multicolumn{1}{l|}{3.81e-07***} & \multicolumn{1}{l|}{2.47e-07***} & \multicolumn{1}{l|}{3.21e-07***} & \multicolumn{1}{l|}{-}           & 8.10e-02    \\ \hline
\multicolumn{1}{|l|}{A-CNN}         & \multicolumn{1}{l|}{1.59e-06***} & \multicolumn{1}{l|}{8.84e-07***} & \multicolumn{1}{l|}{3.03e-06***} & \multicolumn{1}{l|}{8.10e-02}    & -           \\ \hline
\multicolumn{1}{|l|}{SVM}           & \multicolumn{1}{l|}{1.46e-06***} & \multicolumn{1}{l|}{5.15e-07***} & \multicolumn{1}{l|}{1.14e-06***} & \multicolumn{1}{l|}{3.91e-04***} & 1.70e-04*** \\ \hline
\multicolumn{1}{|l|}{FBCCA}         & \multicolumn{1}{l|}{2.69e-06***} & \multicolumn{1}{l|}{3.86e-06***} & \multicolumn{1}{l|}{9.03e-06***} & \multicolumn{1}{l|}{2.33e-02*}   & 1.08e-03*** \\ \hline
\multicolumn{1}{|l|}{Random Forest} & \multicolumn{1}{l|}{2.48e-07***} & \multicolumn{1}{l|}{2.48e-07***} & \multicolumn{1}{l|}{2.48e-07***} & \multicolumn{1}{l|}{2.48e-07***} & 2.70e-07*** \\ \hline
\multicolumn{6}{|c|}{\textbf{BETA Dataset}}                                                                                                                                                   \\ \hline
\multicolumn{1}{|l|}{FBRNN}         & \multicolumn{1}{l|}{-}           & \multicolumn{1}{l|}{5.41e-01}    & \multicolumn{1}{l|}{5.39e-02}    & \multicolumn{1}{l|}{3.62e-11***} & 3.16e-08*** \\ \hline
\multicolumn{1}{|l|}{FBCNN-3D}      & \multicolumn{1}{l|}{5.41e-01}    & \multicolumn{1}{l|}{-}           & \multicolumn{1}{l|}{1.33e-01}    & \multicolumn{1}{l|}{4.60e-12***} & 2.72e-10*** \\ \hline
\multicolumn{1}{|l|}{FBCNN-2D}      & \multicolumn{1}{l|}{5.39e-02}    & \multicolumn{1}{l|}{1.33e-01}    & \multicolumn{1}{l|}{-}           & \multicolumn{1}{l|}{8.22e-12***} & 2.04e-08*** \\ \hline
\multicolumn{1}{|l|}{A-RNN}         & \multicolumn{1}{l|}{3.62e-11***} & \multicolumn{1}{l|}{4.60e-12***} & \multicolumn{1}{l|}{8.22e-12***} & \multicolumn{1}{l|}{-}           & 4.11e-07*** \\ \hline
\multicolumn{1}{|l|}{A-CNN}         & \multicolumn{1}{l|}{3.16e-08***} & \multicolumn{1}{l|}{2.72e-10***} & \multicolumn{1}{l|}{2.04e-08***} & \multicolumn{1}{l|}{4.11e-07***} & -           \\ \hline
\multicolumn{1}{|l|}{SVM}           & \multicolumn{1}{l|}{1.12e-10***} & \multicolumn{1}{l|}{3.01e-10***} & \multicolumn{1}{l|}{1.08e-08***} & \multicolumn{1}{l|}{8.98e-05***} & 1.65e-01    \\ \hline
\multicolumn{1}{|l|}{FBCCA}         & \multicolumn{1}{l|}{2.56e-03***} & \multicolumn{1}{l|}{8.97e-03*}   & \multicolumn{1}{l|}{1.64e-01}    & \multicolumn{1}{l|}{9.79e-09***} & 1.31e-03*** \\ \hline
\multicolumn{1}{|l|}{Random Forest} & \multicolumn{1}{l|}{8.76e-13***} & \multicolumn{1}{l|}{4.31e-13***} & \multicolumn{1}{l|}{7.89e-13***} & \multicolumn{1}{l|}{1.50e-11***} & 2.08e-12*** \\ \hline
\multicolumn{6}{|c|}{\textbf{Portable Dataset}}                                                                                                                                               \\ \hline
\multicolumn{1}{|l|}{FBRNN}         & \multicolumn{1}{l|}{-}           & \multicolumn{1}{l|}{5.86e-03*}   & \multicolumn{1}{l|}{3.91e-03***} & \multicolumn{1}{l|}{2.73e-02*}   & 3.75e-01    \\ \hline
\multicolumn{1}{|l|}{FBCNN-3D}      & \multicolumn{1}{l|}{5.86e-03*}   & \multicolumn{1}{l|}{-}           & \multicolumn{1}{l|}{6.45e-02}    & \multicolumn{1}{l|}{3.91e-03***} & 1.95e-03*** \\ \hline
\multicolumn{1}{|l|}{FBCNN-2D}      & \multicolumn{1}{l|}{3.91e-03***} & \multicolumn{1}{l|}{6.45e-02}    & \multicolumn{1}{l|}{-}           & \multicolumn{1}{l|}{1.95e-03***} & 1.95e-03*** \\ \hline
\multicolumn{1}{|l|}{A-RNN}         & \multicolumn{1}{l|}{2.73e-02*}   & \multicolumn{1}{l|}{3.91e-03***} & \multicolumn{1}{l|}{1.95e-03***} & \multicolumn{1}{l|}{-}           & 9.77e-03*   \\ \hline
\multicolumn{1}{|l|}{A-CNN}         & \multicolumn{1}{l|}{3.22e-01}    & \multicolumn{1}{l|}{1.95e-03***} & \multicolumn{1}{l|}{1.95e-03***} & \multicolumn{1}{l|}{9.77e-03*}   & -           \\ \hline
\multicolumn{1}{|l|}{SVM}           & \multicolumn{1}{l|}{1.37e-02*}   & \multicolumn{1}{l|}{1.95e-03***} & \multicolumn{1}{l|}{1.95e-03***} & \multicolumn{1}{l|}{1.95e-02*}   & 1.95e-03*** \\ \hline
\multicolumn{1}{|l|}{FBCCA}         & \multicolumn{1}{l|}{1.95e-03***} & \multicolumn{1}{l|}{1.95e-03***} & \multicolumn{1}{l|}{1.95e-03***} & \multicolumn{1}{l|}{1.95e-03***} & 1.95e-03*** \\ \hline
\multicolumn{1}{|l|}{Random Forest} & \multicolumn{1}{l|}{1.95e-03***} & \multicolumn{1}{l|}{1.95e-03***} & \multicolumn{1}{l|}{1.95e-03***} & \multicolumn{1}{l|}{5.86e-03*}   & 1.95e-03*** \\ \hline
\end{tabular}
\caption{Paired Wilcoxon signed-rank tests comparing the accuracy of different models, in the three datasets. Cells show p-value. * indicates statistically significant results, with p$\leq$0.05. *** means p$\leq$0.005.}
\label{statistics}
\end{table}

\section{Discussion}
%fb, subject-independent, recurrent vs cnn, worst subjects, fbcca, best performing always fb. time information.

Firstly, we observed that the filter banks had a strong positive impact on DNN performance. Analyzing the mean accuracies and F1-Scores provided by the tests performed in the 3 datasets, we see that the FBRNN always surpassed the A-RNN, the FBCNN-3D was better than the A-CNN, and the FBCNN-2D exceeded the A-CNN and SVM. The tests in table \ref{statistics} support these claims, as the models mean accuracy differences are statistically significant (with p$\leq$0.05 or, mostly, p$\leq$0.005). Taking the Benchmark dataset results as an example, we find the largest improvement in the best performing model, the FBRNN; comparing its mean accuracy and F1-Score, 87.3\% and 0.877, to its counterpart without the filter bank (the A-RNN, with 82.7\% and 0.829), we observe a boost of 4.6\% and 0.048 in accuracy and F1-Score, respectively. A similar phenomenon occurred with the CNNs: comparing the two similar networks, FCBNN-3D and A-CNN, it is noteworthy that the filter bank increased mean accuracy by 3.6\% and mean F1-Score by 0.041, from 83.1\% and 0.831 in the A-CNN to 86.7\% and 0.872 in the FBCNN-3D. The superior performances of FBDNNs makes the benefits of filter banks clear. In the cross-dataset testing (table \ref{BetaTest}) we see that the FBDNNs could still achieve mean accuracies close to 80\% (table \ref{BetaTest}), indicating strong generalization capability.

Comparing the recurrent networks to the convolutional ones, considering the DNNs without filter banks, table \ref{res}, \ref{BetaTest} and \ref{portable} show a superior result for the convolutional model. Table \ref{statistics} indicates that these accuracy differences are statistically significant in the BETA and Portable datasets, while the Benchmark dataset showed a p-value of 0.081. With the filter bank the recurrent model shows a slightly superior performance on the Benchmark and BETA datasets, with a statistically significant difference only on Benchmark (p$\leq$0.05). But, in the Portable dataset (table \ref{portable}), the FBCNN-3D surpasses the FBRNN, with 4.9\% better mean accuracy (significantly differing, p$\leq$0.05). The recurrent neural networks access the original time signal, without any loss of information caused by FFT/STFT. When this signal is cleaner (i.e., created by a medical grade, stationary EEG system, as in the Benchmark and BETA datasets), the FBRNN and A-RNN show strong performances. However, the recurrent networks seem more sensitive to noise, as their performances deteriorate in the wearable BCI, whose interface does not have the same quality and SNR as the stationary systems.

The weak performances obtained in the Portable dataset reinforces the idea that high accuracy is still not achievable with the tested models in such an extreme scenario (12 targets, 0.5 s data length, a single electrode and a calibration-free BCI). The employed data length limits the frequency resolution in the FFT/STFT to 2 Hz. Thus, having stimulus frequencies separated by 0.5 Hz, the models employing the transformations must rely more on the SSVEP harmonics for classification. The higher the harmonic, the lower the SNR in SSVEP \cite{FBCCA}, reducing classification accuracy. Indeed, the authors are not aware of models that can accurately solve this SSVEP classification task. However, table \ref{portable} shows that FBDNNs represent an improvement in relation to the state-of-the-art, due to its ability to better extract information from the harmonics. 

Even being relatively small and easy to run (in the context of DNNs), all the deep neural networks in this study clearly surpassed the shallow and linear SVM and the Random Forest. Probable reasons for this result are that the DNNs are more flexible models and analyze more informative inputs. While the SVM and Random Forest input (magnitude spectrum) contains only information related to the power spectrum, the FCBNN-2D analyzes complex spectrum features, containing phase and magnitude information. As a previous article \cite{ComplexSSVEP} has shown, the additional phase information in complex inputs improves classification performances, particularly when different SSVEP visual stimuli present distinct phases. Furthermore, in this study, the FBCNN-3D and A-CNN performances confirmed the usefulness of an even more information rich input format, the complex spectrogram, which additionally presents the time dimension. In the three datasets the 3D network scores among the best performing models, being the best performer in the Portable database. It also surpassed the 2D model in every database, even though the accuracy difference was small (with a minimum p-value of 0.0645, in the Portable dataset). Finally, the RNNs analyze the original time signal, which did not lose information with STFT or FFT transformations. 

However, we must note that the inferior FBCCA accuracy in this configuration is an expected result, because it was conceived as a multi-channel technique and it loses much performance when working with a single electrode. The only case in which FBCCA matches or surpasses some of the neural networks is when they were trained in the Benchmark dataset and tested on BETA, because FBCCA does not require training. Even in this case, the FBDNNs excel it.

All the proposed models are small and fast during run-time. Thus, their size would not be a problem for online applications. Table \ref{perf} shows the necessary time to classify a signal, and the number of model parameters, for the methods that we implemented using the graphics processing unit (NVidia RTX 3080). Note that the methods consider a recorded signal of 0.5 s. Furthermore, the filter banks are applied to the data stream (before window slicing), which shall have a longer duration, for them to produce an adequate response. The Random Forest and FBCCA were implemented with the CPU (AMD Ryzen 9 5900X). They take 4.18 ms and 7.82 ms to classify a sample, respectively. We observe that, in the proposed configurations, the neural networks with filter banks have a remarkably similar run-time speed in relation to their alternative counterparts. Training time was also not strongly affected when using FBDNNs. For example, in the Portable dataset one epoch took about 4.8 s for the A-CNN, and 5.5 s for the FBCNN-3D; the FBRNN and the A-RNN had both epochs of about 12 s.

\begin{table}[]
\centering
\begin{tabular}{|l|l|l|}
\hline
Method   & \begin{tabular}[c]{@{}l@{}}Classification\\ Time (ms)\end{tabular} & Parameters \\ \hline
FBRNN    & 1.011                                                              & 87836      \\ \hline
FBCNN-3D & 0.236                                                              & 23378      \\ \hline
FBCNN-2D & 0.224                                                              & 6674       \\ \hline
A-CNN    & 0.233                                                              & 6290       \\ \hline
A-RNN    & 1.009                                                              & 85532      \\ \hline
SVM      & 0.034                                                              & 46         \\ \hline
\end{tabular}
\caption{Model run-time performance and size.}
\label{perf}
\end{table}

\section{Conclusion}
%complex spec, rnn vs cnn

The filter banks allowed the deep neural networks to more efficiently analyze the harmonic components of SSVEP, improving classification performance in a single-channel and user-independent (also known as cross-subject or calibration-free) BCI with small data length (0.5 s). We perceived this effect in a recurrent neural network, in a CNN analyzing complex spectrum features, and in a CNN processing complex spectrograms. According the results achieved in the 3 employed datasets, the proposed FBDNNs surpass their counterparts without filter banks. This study also showed that FBDNNs strongly outperformed classical SSVEP classification methods in the proposed BCIs. For example, in the Benchmark dataset, the FBRNN mean test accuracy was 6.4\% and 7.1\% higher than the SVM's and FBCCA's, respectively.

The analysis of complex spectrum features, introduced by \cite{ComplexSSVEP}, improved SSVEP classification accuracy in relation to classifying the magnitude spectrum, because it creates DNN inputs that carry magnitude and phase information. The complex spectrogram, introduced in this study, creates an even more information rich input by adding the time dimension. Processing these objects with a 3D CNN (FBCNN-3D) demonstrated strong performances, and produced the most capable model for the Portable dataset. These results point out that DNNs can efficiently analyze more detailed inputs, which carry more information, thus improving performance in SSVEP classification.

Even though they are deep, the FBDNNs proposed in this study do not contain a very large number of parameters and are adequate for online applications and to be implemented with less powerful electronic devices. The trained DNNs will be available for download (https://github.com/PedroRASB/FBCNN), and we hope they can help future studies in this field. 

Finally, we must state that the obtained mean accuracy and F1-Score are very positive results, because the addressed classification problem is not simple. Firstly, using a single electrode reduces SSVEP classification accuracy, as does the small data length \cite{singleChenel2}, \cite{SingleChannelCNN}. Furthermore, subject-independent BCIs are known to be more challenging and prone to overfitting \cite{ComplexSSVEP}. However, improving BCI performance in this complicated scenario is essential for the construction of devices that are faster, have lower latency and are more portable and economical.

\section{Acknowledgments}
This work was partially supported by CNPq (process 308811/2019-4) and CAPES.
The authors declare no competing interests.

\bibliographystyle{abbrv}
\bibliography{mybibfile}

\end{document}